\documentclass[a4paper,twocolumn]{spaceDebrisC}
\usepackage{times}
\usepackage{graphicx}
\usepackage[square,numbers]{natbib}
\usepackage{url}
\usepackage{hyperref}

\title{Understanding the impact of satellites on radio astronomy observations}
\author{M. Peel}
\affil{Imperial College London, Blackett Lab, Prince Consort Road, London SW7 2AZ, UK, m.peel@imperial.ac.uk}
\author{S. Eggl}
\affil{Department of Aerospace Engineering/Department of Astronomy, University of Illinois at Urbana-Champaign, Illinois, USA.}
\author{M. L. Rawls}
\affil{Department of Astronomy, University of Washington, Seattle, WA, USA}
\author{H. Qiu}
\affil{SKA Observatory, Jodrell Bank, Macclesfield SK11 9FT, United Kingdom.}
\author{D. L. Clements}
\affil{Imperial College London, Blackett Lab, Prince Consort Road, London SW7 2AZ, UK}

\begin{document}


\maketitle

\begin{abstract}
Radio telescopes observe extremely faint emission from astronomical objects, ranging from compact sources to large scale structures that can be seen across the whole sky. Satellites actively transmit at radio frequencies (particularly at 10--20\,GHz, but usage of increasing broader frequency ranges are already planned for the future by satellite operators), and can appear as bright as the Sun in radio astronomy observations. Remote locations have historically enabled telescopes to avoid most interference, however this is no longer the case with dramatically increasing numbers of satellites that transmit everywhere on Earth. Even more remote locations such as the far side of the Moon may provide new radio astronomy observation opportunities, but only if they are protected from satellite transmissions. Improving our understanding of satellite transmissions on radio telescopes across the whole spectrum and beyond is urgently needed to overcome this new observational challenge, as part of ensuring the future access to dark and quiet skies.

In this contribution we summarise the current status of observations of active satellites at radio frequencies, the implications for future astronomical observations, and the longer-term consequences of an increasing number of active satellites. This will include frequencies where satellites actively transmit, where they unintentionally also transmit, and considerations about thermal emission and other unintended emissions. This work is ongoing through the IAU CPS.
\end{abstract}

\section{Radio astronomy observations}

Radio astronomy relies on a wide variety of antennas and receiver packages to study the radio sky at various resolutions and frequencies. It largely operates passively, coexisting with other users of the radio spectrum. A few small frequency bands are reserved by the Radiocommunication sector of the International Telecommunication Union (ITU-R) for Radio Astronomy Services (RAS), in particular for observations of spectral lines and physical processes that occur at certain fixed frequencies.However, for most observations these narrow frequency bands are not sufficient to achieve the required sensitivities on the sky. To observe very faint astronomical observations radio telescopes operate with broadband receivers across a wide range of frequencies.

This approach necessitates avoiding or removing the emission generated by other users of the radio spectrum. Artificial signals are generally many orders of magnitude brighter than the astronomical signals. A first step to improve the signal-to-noise ratio is to observe from remote regions where there are few people and sources of radio transmissions, for example in the Karoo desert in South Africa and Murchison region in Australia for the Square Kilometer Array (SKA), or from the Atacama plateau in Chile for the Atacama Large Millimeter Array (ALMA) and Simons Observatory. These can either have radio quiet zones defined around them: giving additional legal protection to limit ground-based transmissions, or rely on their remoteness to avoid significant human-made transmissions. This has worked well for many years, however the situation has changed dramatically due to the recent advent of satellite constellations, particularly in Low Earth Orbit (LEO), which aim to have satellite coverage everywhere on Earth and specifically try to fill in missing coverage by ground-based transmitters in remote locations (where it is affordable \cite{rawls_satellite_2020}). Local radio quiet zones do not legally prevent non-ground-based transmissions, although some operators avoid transmissions towards such sites where possible.

Understanding the radio frequency (RF) properties of satellite constellations is thus now very important to continue the operations of these remote astronomy facilities, however not much information is currently available from operators, so this relies on measurements by the observatories themselves. This is enabled through groups like the International Astronomical Union's Centre for the Protection of the Dark and Quiet Sky from Satellite Constellation Interference (IAU CPS)\footnote{\url{https://cps.iau.org/}} \cite{2024Peel,2024Rawls}, Committee on Radio Astronomy Frequencies (CRAF) \cite{2023Winkel}, and similar radio astronomy related organisations around the world, along with a series of conferences including SATCON 1 \cite{walker_impact_2020} and 2 \cite{hall_executive_2021,rawls_satcon2_2021}, and the Dark and Quiet Skies conference 1 \cite{iau_dark_2021} and 2 \cite{walker_dark_2022}.

In this proceeding we summarise the current understanding of satellite emissions at radio frequencies through their active transmissions  in Section 2 and their unintentional emissions, both radio and thermal, in Section 3, before concluding in Section 4.

\section{Active transmissions}
\begin{table}
    \centering
    \caption{Some of the active transmission frequencies used by satellite constellations that can affect radio astronomy observations.}\vspace{1em}
    \renewcommand{\arraystretch}{1.2}
    \begin{tabular}{lll}
        \hline
        \textbf{Satellites} & \textbf{Use} & \textbf{Frequencies} \\
        \hline
        Starlink & DTC downlink & 1.190--1.995\,GHz\\ 
        & User downlink & 10.7--12.7\,GHz\\
         & Gateway downlink & 17.8--18.6\,GHz\\
         & Gateway downlink & 18.8--19.3\,GHz\\
         & Gateway downlink & 19.7--20.2\,GHz\\
         & User downlink & 37.5--42.5\,GHz\\
         & Gateway downlink & 37.5--37.75\,GHz\\
         \hline
        OneWeb & User downlink & 10.7--12.7\,GHz\\
         & Gateway downlink & 17.8--18.6\,GHz\\
         & Gateway downlink & 18.8-19.3\,GHz\\
         \hline
        Amazon Kuiper & User/GW downlink & 17.7--18.6\,GHz\\
         & User/GW downlink & 18.8--19.3\,GHz\\
         & User/GW downlink & 19.3--19.4\,GHz\\
         & User/GW downlink & 19.7--20.2\,GHz\\
         \hline
    \end{tabular}
    \label{tab:frequencies}
\end{table}

Satellite constellations primarily operate in the frequency range 10--20\,GHz, although there have also been requests to use frequencies around 40\,GHz, and even 120--170\,GHz\footnote{\url{https://www.itu.int/ITU-R/space/asreceived/Publication/DisplayPublication/53068}}. They have also recently expanded to offer direct-to-cell (DTC) communications around 2\,GHz. A short (and incomplete) summary of the relevant frequencies for radio astronomy is given in Table \ref{tab:frequencies}.

\begin{figure}
    \centering
    \includegraphics[width=0.48\linewidth]{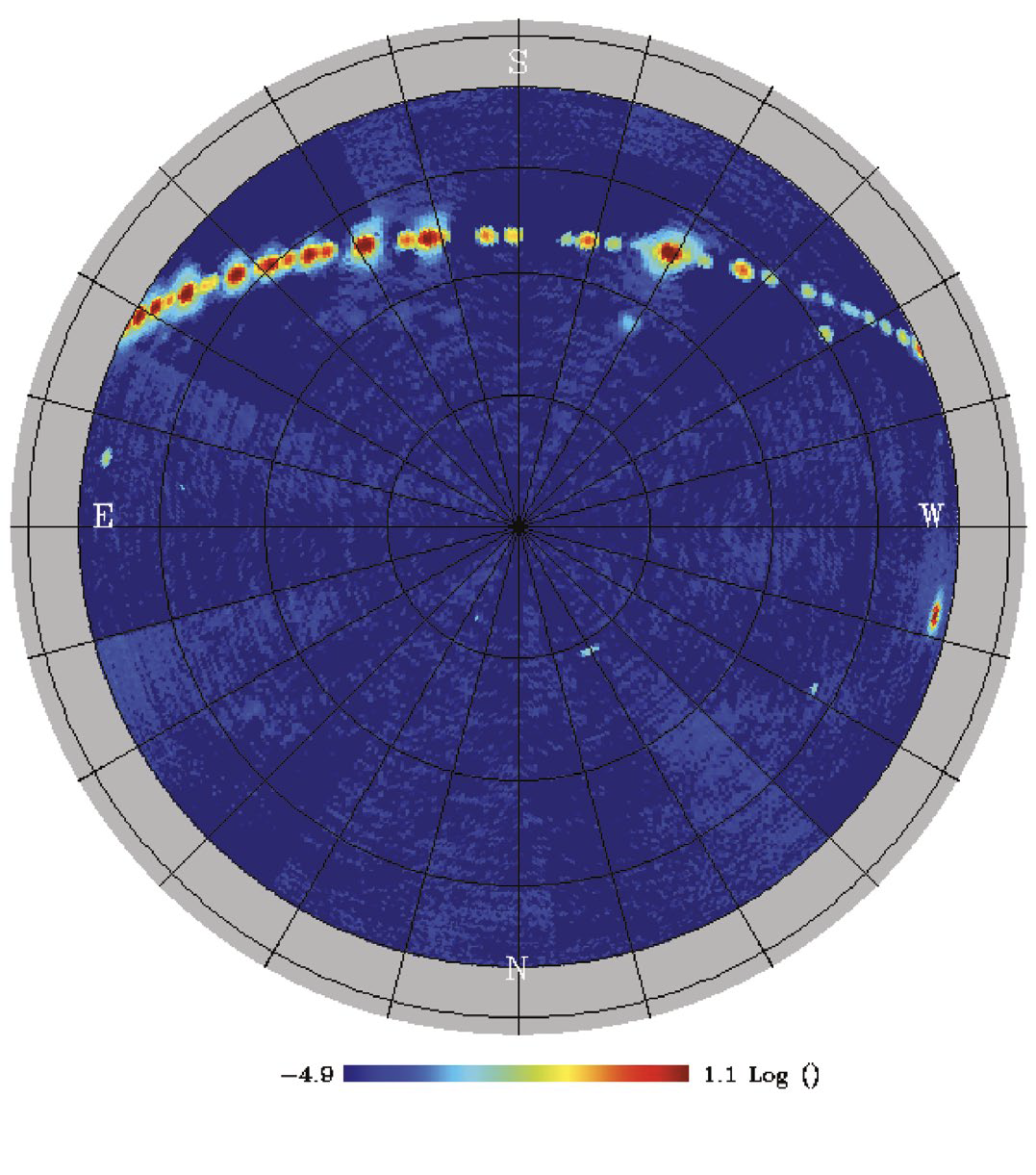}
    \includegraphics[width=0.48\linewidth]{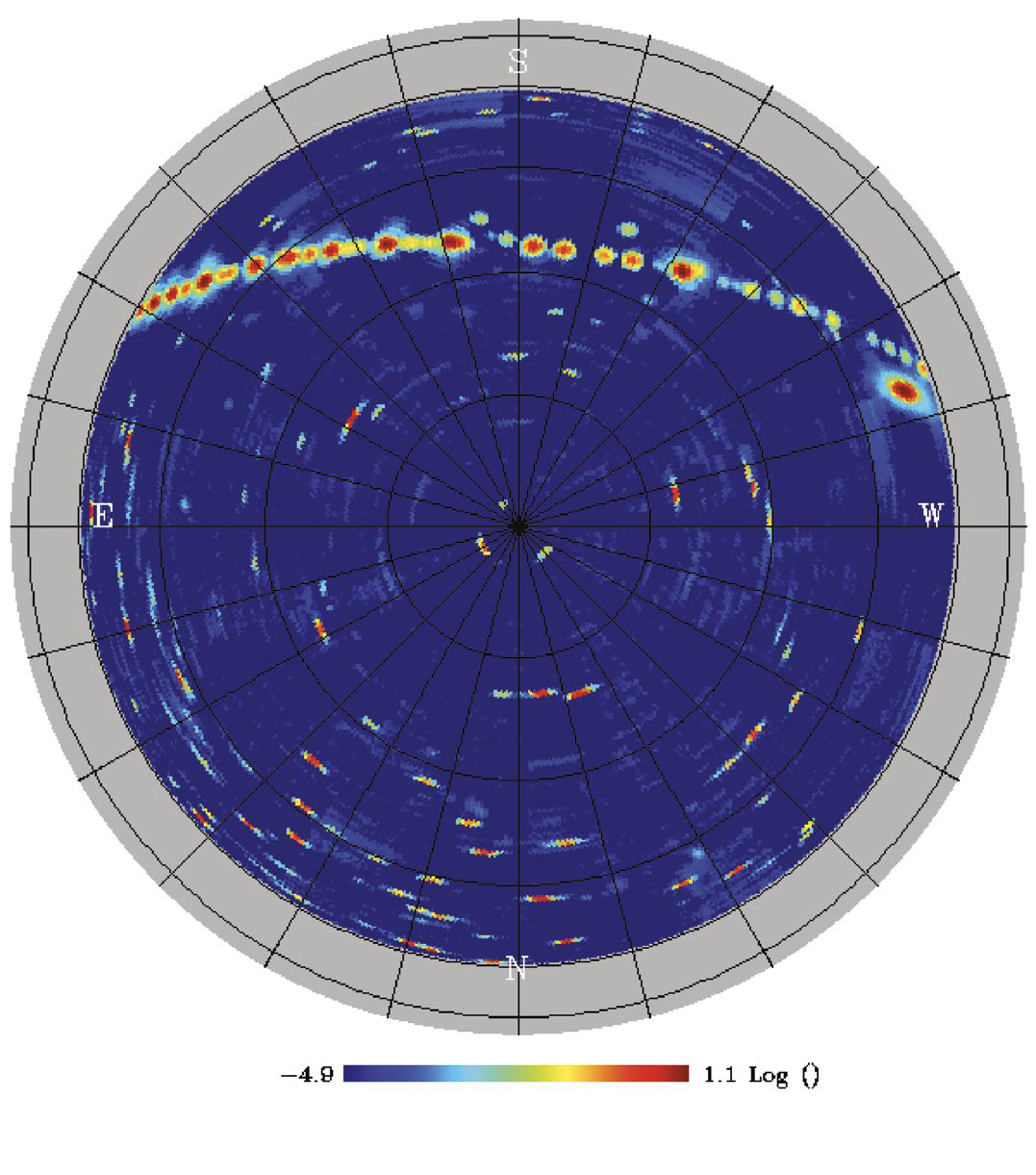}
    \caption{Local sky maps at 10--14\,GHz by QUIJOTE MFI \cite{2023QUIJOTE}. Left: observations from 2014, showing geostationary satellites in a band. Right: observations in 2024 with QUIJOTE MFI2 showing transmissions from satellite constellations (red dots) across the whole sky. Credit: QUIJOTE Collaboration. }
    \label{fig:quijote}
\end{figure}

One telescope that is particularly affected by constellation transmissions is Q-U-I JOint TEnerife (QUIJOTE), which observes at 10--20\,GHz as well as at 30, 40, and 90\,GHz, from Teide Observatory in Tenerife. It uses its remote location and high altitude to avoid emissions, without a formal radio quiet zone. QUIJOTE recently completed its initial survey at 10--20\,GHz with its multi-frequency instrument (MFI) \cite{2023QUIJOTE}, where the main issues caused by satellites were those in geostationary orbit that could appear as bright as the Sun, making part of the sky unobservable at 10--14\,GHz in particular. Starlink, OneWeb and other satellite constellations were identified as a concern for its upcoming MFI2 survey\footnote{\url{http://research.iac.es/proyecto/cmb/pages/posts/impact-of-satellite-ldquomega-constellationsrdquo-on-cmb-experiments-at-the-teide-observatory2.php}}, and initial data has shown that it sees many Starlink and OneWeb satellites across the whole of the sky, and particularly at lower elevation, as shown in Figure~\ref{fig:quijote}. As well as the direct observations of the satellites, the impact that they could have in the telescope sidelobes is also a concern, since that can generate spurious large angular scale emission that could be confused with faint Galactic emission on large scales.

As more satellites are launched in the future, this situation will become increasingly complicated, and a new field-programmable gate array (FPGA) back-end is planned to remove frequencies affected by satellites from QUIJOTE MFI2 observations, using as much of the remaining clear bandwidth as possible to complete its survey \cite{2022Hoyland}. Transmissions at these frequencies will also affect 10--15\,GHz observations by the higher frequency part of the SKA, as well as the Very Large Array (VLA), and a wide variety of large single dish telescopes.

One thing that will likely help is to restrict the use of user terminals near to observatories, which is often but not always possible---particularly with mobile terminals and if other non-radio-frequency scientific instruments nearby choose to use satellite internet. Starlink's available coverage map\footnote{\url{https://www.starlink.com/map}} clearly shows key radio telescope observatories through regions where Starlink is unavailable, although these are not included around all radio telescopes.

As satellites move transmissions to higher frequencies, other facilities could also become increasingly affected. Other cosmic microwave background experiments typically observe from $\sim$30\,GHz upwards, as does ALMA, with other facilities such as AtLAST \cite{2025AtLAST} planned for the future. Ground stations, which typically use higher frequencies from user terminals, have to be located far away from these observatories to avoid potential illumination of these astronomical facilities.

In order to make satellite operators more aware of the locations of radio telescopes that need protection, CPS is currently assembling a list of radio telescopes \cite{2024Sorokin}. This aims to complement approaches like the ITU-R list of radio quiet zones, to cover all telescopes that may or may not have formal protection but could benefit from avoiding being directly illuminated by satellites.

A promising approach to help reduce the impact of satellites on some radio astronomy facilities is to avoid transmissions towards the boresights of telescopes \cite{2024Nhan}. This requires satellite operators knowing which direction radio telescopes are pointing in, so they can turn off the satellite transmitters when they would be passing through the telescope beam. This avoids transmissions towards the main beams of radio telescopes, which will significantly help observations by telescopes involved in this process. However, it is unclear which telescopes, particularly those outside the US, would be able to participate in this. It would not change the satellite emission seen in telescope sidelobes unless satellites are turned off while they are essentially above the horizon for the telescope. It also does not help with telescopes such as SKA-low that use arrays of dipoles that see everywhere on the sky at once, using software to define which direction the telescope is effectively focused towards.

Space radar can also present significant problems for radio astronomy due to their use of very bright transmissions. One notable example of this is CloudSat, which used a 94\,GHz earth-facing radar that could be seen brightly by radio observatories, and was a major concern for ALMA\footnote{\url{https://science.nrao.edu/facilities/alma/aboutALMA/Technology/ALMA_Memo_Series/alma504/memo504.pdf}}. CloudSat is currently being de-orbited, and is being replaced by EarthCARE, which turns off its radar systems when passing over radio observatories\footnote{\url{http://www.iucaf.org/Coordination_Agreement_Between_EARTHCARE_and_IUCAF_Observatories_v2-January_2022.pdf}}. Similar arrangements with others using space-based radar to avoid radio astronomy sites would be beneficial. Radar reflections from space debris illuminating radio telescopes could also be a concern in the future, particularly considering the increasing use of radar to track LEO objects. The use of narrow-frequency satellite beacons that transmit trackable signals has not caused significant issues with radio astronomy, however, particularly due to their narrow bandwidth.\footnote{They are also useful for independent tracking by third parties, e.g., \url{https://satnogs.org/}.}

\section{Unintentional emissions}
\subsection{Low frequencies}
At low ($\sim$100\,MHz) frequencies, all electronics emit radiation due to electronic noise, switching mode power supplies, internal clocks, etc. This is also true for active satellites, and LOFAR and SKA-low prototypes have already detected Starlink satellites through this Unintended ElectroMagnetic Radiation (UEMR) at 100--200\,MHz
\cite{di_vruno_unintended_2023,grigg_detection_2023}, with increasing amounts in the latest generation of Starlink satellites compared to earlier ones \cite{bassa_bright_2024}. This is particularly a concern as satellite numbers increase dramatically, and it can in principle be prevented with improved RF shielding of the electronics on board the satellite bodies.

\subsection{Out-of-band emissions}

Radio systems will also naturally transmit at octaves of active frequencies, and RF filters are never perfect so low-level emission can also be transmitted at adjacent frequencies to the active band. One particular case here is the 10.6--10.7\,GHz RAS protected band, which is directly adjacent to the permitted bands for Starlink and OneWeb. To minimise emissions in this frequency range, the satellite operators voluntarily avoid using their lowest 250\,MHz frequency band. Similar precautions may be needed around other RAS protected bands.

One particular example here is the Iridium satellite constellation uses transmissions at 1.6\,GHz that are directly adjacent to a RAS protected band reserved to observe hydroxyl (OH) emission. Iridium satellites also transmit a comb of signals into that RAS protected band, which can be seen when a satellite passes near to the beam of a radio telescope observing at these frequencies, leading to interference that has to be removed from the RAS observations \cite{2019Deshpande}.

\subsection{Thermal emission}
Objects in orbit also emit thermal radiation depending on their size and temperature. The South Pole Telescope (SPT) \cite{foster_detection_2024} has detected a variety of objects in orbit at 95, 150, and 220\,GHz, with emission consistent with 300K black-body radiation, including Starlink satellites, and an Indian Space Research Organisation (ISRO) Mark III launch vehicle (LVM3) rocket body. Observatories located closer to the equator, such as those in Chile and Hawaii, will see many more orbiting objects through their thermal emission (Peel et al. in prep.).

\subsection{Conclusions}
Radio astronomy observes very faint astronomical sources that can easily be drowned out by human-made transmissions, even unintentionally. Remote sites have historically avoided most ground-based emissions, but satellites now transmit from orbit everywhere on Earth in rapidly increasing numbers. Avoiding satellite transmissions towards existing and future radio astronomy stations helps mitigate these issues.

The rise of satellite constellations (and the accompanying dramatic reduction in launch costs) has prompted renewed discussions about observing from more remote locations---in space, or potentially on the far side of the Moon, which is naturally sheltered from radio emissions from Earth (and can also be used to observe at frequencies that are normally blocked by the Earth's atmosphere). While it costs significantly more to operate in these environments, and it is not easy to fix any issues that occur with space telescopes, this may become necessary as the Earth orbit becomes more radio loud. However, care should be taken to avoid using the same frequencies onboard satellites in lunar orbit, and also in more remote locations where radio telescopes could be located in the future.

\section*{Acknowledgments}
M.P., S.E., and M.L.R. are co-leads of SatHub, part of the International Astronomical Union's Centre for the Protection of the Dark and Quiet Sky from Satellite Constellation Interference.
S.E. acknowledges support from the National Science Foundation through the following awards: Collaborative Research: SWIFT-SAT: Minimizing Science Impact on LSST and Observatories Worldwide through Accurate Predictions of Satellite Position and Optical Brightness NSF Award Number: 2332736. 

\bibliography{refs} 

\end{document}